\documentclass[useAMS,usenatbib]{mn2e}
\usepackage{txfonts}
\usepackage{longtable}
\usepackage{rotating}
\usepackage{graphicx}
\usepackage{graphics}
\usepackage{psfrag}
\usepackage{amssymb}

\bibpunct[, ]{}{}{;}{a}{}{,}

\def\aap{A\&A}
\def\apj{ApJ}

\def\mnras{MNRAS}

\def\aaps{A\&A Supp.}
\def\apss{Ap\&SS}      
\def\apjs{ApJS}

\newcommand{\teff}{$T_{\rm eff}$}

\newcommand{\lgg}{$\log g$}
\def\vt{$\xi_{\rm t}$}
\newcommand{\ion}[2]{#1\,{\sc #2}}

\newcommand{\ms}{m\,s$^{-1}$}
\newcommand{\vs}{$v_{\rm e}\sin\,i$}
\newcommand{\kms}{km\,s$^{-1}$}

\newcommand{\bz}{$\langle B_{\rm z} \rangle$}

\def\atlas{{\sc ATLAS9}}
\def\width{{\sc WIDTH9}}
\def\synth{{\sc Synth3}}
\def\vald{{\sc VALD}}

\title[Time-resolved photometric and spectroscopic analysis of a luminous Ap star HD\,103498]
{Time-resolved photometric and spectroscopic analysis of a luminous Ap star HD\,103498%
\thanks{The spectroscopic observations are made from the Nordic Optical Telescope operated on the island of La Palma jointly by Denmark, Finland, Iceland, Norway and Sweden, in the Spanish Observatorio del Roque de los Muchachos of the Instituto de Astrofisica de Canarias.}}
\author[S. Joshi et al.]
{S. Joshi$^1$\thanks{E-mail:santosh@aries.res.in},
T. Ryabchikova$^{2,3}$, O. Kochukhov$^{4}$, M. Sachkov$^{2}$, S. K. Tiwari$^1$, \and N. K. Chakradhari$^{5}$, N. Piskunov$^{4}$ \\
$^{1}$ Aryabhatta Research Institute of Observational Sciences (ARIES), Manora Peak, Nainital, India \\
$^{2}$ Institute of Astronomy, Russian Academy of Sciences, Pyatnitskaya 48, 119017 Moscow, Russia \\
$^{3}$	Institut f\"ur Astronomie, Universit\"{a}t Wien, T\"{u}rkenschanzstrasse 17, 1180 Wien, Austria\\
$^{4}$ Department of Physics and Astronomy, Uppsala University, SE-751 20, Uppsala, Sweden \\
$^{5}$ School of Studies in Physics and Astrophysics, Pt. Ravishankar Shukla University, Raipur, India \\
}

\begin{document}

\maketitle

\begin{abstract}
We present the results on the photometric and spectroscopic monitoring of a luminous Ap star HD\,103498.
The time-series photometric observations were carried out on 17 nights  using three-channel fast photometer attached to the
1.04-m optical telescope at  ARIES, Nainital. The photometric data of five nights of year 2007 show clear signature of 15-min periodicity.
However, the follow-up observations during 2007--2009 could not repeated any such periodicity. To confirm the photometric light
variations, the time-series spectroscopic observations were carried out with the 2.56-m Nordic Optical Telescope
(NOT) at La Palma on February 2, 2009. Any radial velocity variations were absent in this data set which is in full agreement with
the photometric observations taken near the same night. Model atmosphere and abundance analysis of HD\,103498 show that the star is
evolved from the Main Sequence and its atmospheric abundances are similar to two other evolved Ap stars HD\,133792 and HD\,204411:
large overabundances of Si, Cr, and Fe and moderate overabundances of the rare-earth elements. These chemical properties and a higher effective temperature distinguish HD\,103498 from any known roAp star.
\end{abstract}


\section{Introduction}
\label{introduction}
The chemically peculiar (CP) star HD\,103498 (65\,UMa\,D, HR\,4561) is a member of a multiple-system which consists of four objects.
According to Pourbaix et al. (\cite{tokovinin}), HD\,103498 is a south-eastern component of the visual binary ADS\,8347 (separation 63$^{\prime\prime}$) containing HR\,4561 and HR\,4560 with the latter being a spectroscopic binary  65\,UMa\,AC.
Abt \& Morrell (\cite{AM95}) classified it as CrSrEu based on low-resolution Cassegrain photographic spectrum.
The first magnetic field measurements of 65\,UMa\,D were reported by Bychkov et al. (\cite{bychkov03}) who discovered a rather strong negative longitudinal magnetic field \bz\,  of about $-800$~G. Using the MuSiCoS spectropolarimeter attached to 2.0-m
 Telescope Bernard Lyot (TBL) at Observatoire du Pic du Midi, France,
Auri\`ere~et~al. (\cite{AWS07}) measured much weaker longitudinal field
\bz, that varied between $\pm$200~G with a period of  15.83 days. The authors derived rotational velocity \vs=13~\kms\, and estimated
an effective temperature \teff=9220$\pm$300~K and luminosity $\log(L/L_{\odot})$=2.06$\pm$0.20 of the star.

\begin{figure*}
\hbox{
\includegraphics[width=8.5cm,height=12.0cm]{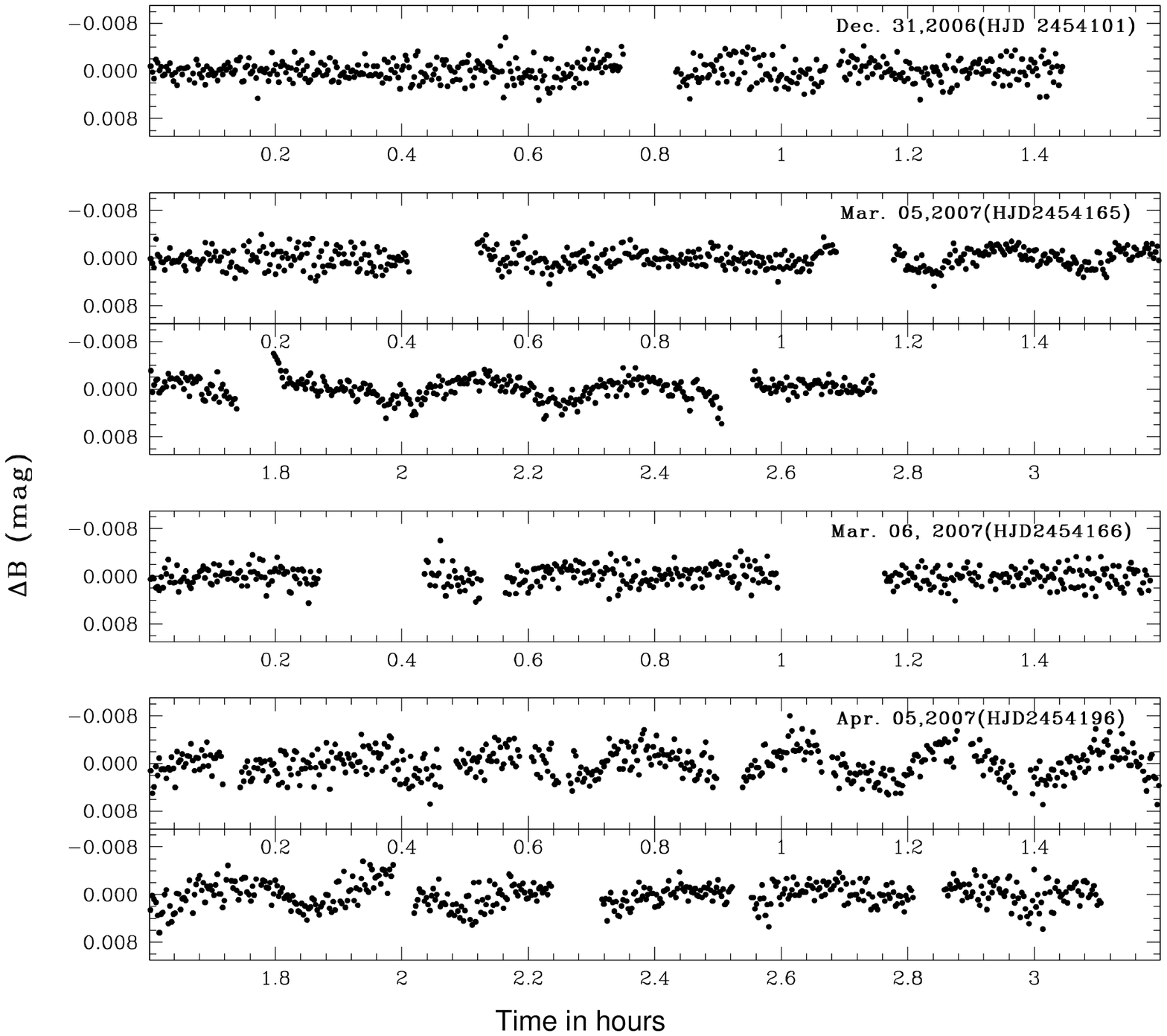}
\includegraphics[width=8.5cm,height=12.0cm]{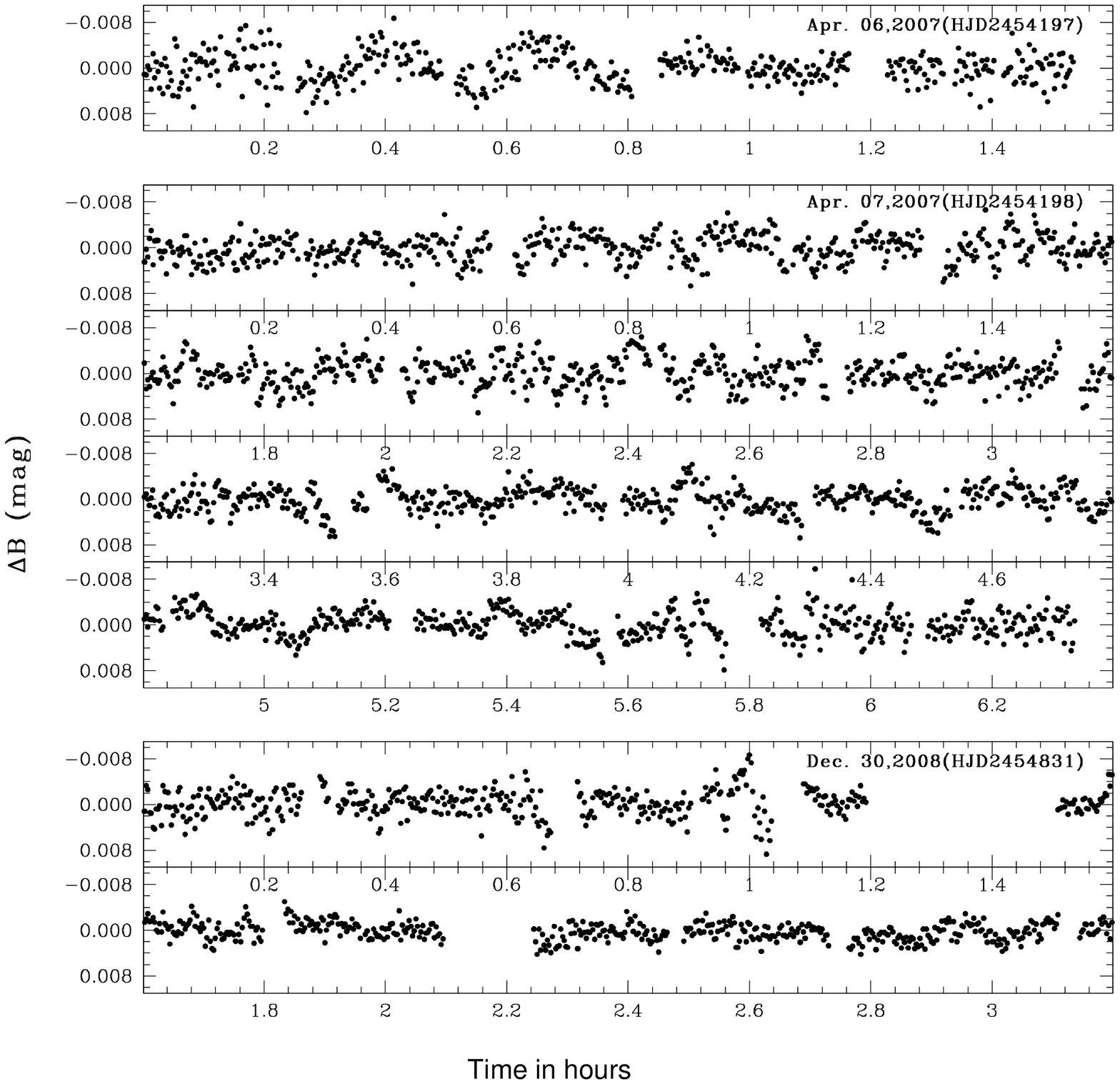}
}
\caption{Light curves of HD\,103498 obtained on different nights between year 2006 to 2008. The HJD (2450000+) of the each observation is mentioned in each panels.}
\label{light}
\end{figure*}

The Str\"{o}mgren indices of HD\,103498 are $b-y$ = 0.003, $m_1$ = 0.196, $c_1$ = 1.010 and $H_{\beta}$ = 2.858
(Hauck \& Mermilliod \cite{Hauck_M}).
On the basis of the peculiar indices and Ap spectral classification, HD\,103498 was observed using high-speed photometric technique
as a part of survey programme for searching photometric variability in the CP stars. Tiwari et al. (\cite{tiwari07})
reported a 15-min periodic oscillation with varying amplitude from night to night. The location of this star in the H-R diagram (see
Fig. \ref{HR}) towards the larger luminosity and higher temperature than any roAp star observed till now,
makes it an unusual pulsator.  Therefore, we continued  further photometric monitoring of this star. In addition, we obtained  the time-series spectroscopy to search for possible rapid radial velocity (RV) variations because in almost all known roAp stars, the RV amplitudes exceed the photometric amplitudes  significantly  and the pulsational signal may be detected in RV with no detection in the photometry (Hatzes \& Mkrtichian \cite{HM04} -- $\beta$CrB; Kochukhov et al. \cite{HD75445} -- HD\,75445).

This paper presents the combined results from both the photometric and spectroscopic observations of HD\,103498 and is organized as follows: The photometric observations and data analysis are described in Sec. 2. The results obtained from the high-resolution spectroscopy  are presented in Sec. 3.
In Sec. 4 we have discussed the implications of variability and the conclusions drawn from these observations are outlined in Sec. 5.

\section{Photometric Observations and Data Analysis}

To search the photometric light variations in the short period ($\sim$ min) pulsating variables, high-speed photometry is an established technique and is being used since 1980s.  The high-speed photometric observations of HD\,103498 was carried out using a three-channel fast photometer attached to the 1.04-m Sampurnanand telescope of ARIES which is situated  at altitude $\sim$1950-m above the sea level with longitude of 79$^\circ 27'$ E and   latitude 29$^\circ 22'$ N having average FWHM seeing of the order of 2$^{\prime\prime}$  (Ashoka et al.  \cite{ashoka01}; Sagar \cite{sagar99}).  We obtained time-series photometric data through a Johnson $B$-filter with an integrations of 10-sec each for a  total duration of 36-hrs  (Table \ref{log}). An aperture of 30$^{\prime\prime}$ was used to minimize flux variations caused by seeing fluctuations and guiding. The centering of the star  was checked  using manual guiding in a non-periodic time-interval.  The data reduction processes involves the following steps : (a) visual inspection of the light curve to identify and remove the bad data points; (b)  correction for coincident counting losses; (c) subtraction of the interpolated sky background and (d) correction for the mean atmospheric extinction.  After applying these corrections, the times of the mid-points of each data points are converted into heliocentric Julian dates (HJD) with an accuracy of $10^{-5}$ day ($\sim$1-sec). The reduced data comprised of a time-series of the HJD and $\Delta$B magnitudes with respect to the mean values of the run (light curve). The frequency analysis was performed using  Deeming's Discrete Fourier Transform (DFT) for unequally spaced data (Deeming \cite{deeming75}). The frequency analysis produces an amplitude spectrum which provides the frequencies, amplitude and phase that act as an input parameters for the further analysis.

\begin{table}
\caption{Journal of the photometric observations of HD\,103498 carried out from ARIES.
The HJD is 2450000+ and  ``-'' in the frequency and amplitude column indicate no light variations.
The magnetic phase $\phi$ corresponds to $t_0$=HJD\,2450000, z is  signal to noise ratio
in power and F is the false alarm probability.}
\label{log}
\begin{scriptsize}
\begin{tabular}{|r|c|c|c|c|c|c|c|}
\hline
Set &  Start time & $\Delta$t & $f$      & $A$    & $\phi$    & z & F \\
No. & (HJD)       & (hr)      & (mHz)~~~ &(mmag)  & ($\pm$0.2)&   & \\
\hline
1. &  4101.46794 & 1.44 & -                & -    & 0.10     &       &        \\
2. &  4165.20987 & 2.74 & 1.06 $\pm $ 0.10 & 0.95 & 0.12     & 22.56 &7.84E-8 \\
3. &  4166.19829 & 1.85 & 1.10 $\pm$ 0.13  & 0.49 & 0.19     & 9.38  & 0.0277 \\
4. &  4196.15852 & 3.10 & 1.18 $\pm$ 0.12  & 1.69 & 0.08     & 71.40 & $\sim 0$ \\
5. &  4197.12757 & 1.82 & 1.09 $\pm$ 0.17  & 1.71 & 0.14     & 73.10 & $\sim 0$ \\
6. &  4198.10264 & 6.33 & 1.13 $\pm$ 0.03  & 0.81 & 0.21     & 16.40 & 8.59E-5\\
7. &  4455.45086 & 1.03 & -                & -    & 0.46     &       &           \\
8. &  4459.46167 & 1.35 & -                & -    & 0.71     &       &           \\
9. &  4584.14787 & 1.22 & -                & -    & 0.59     &       &           \\
10.&  4585.11297 & 2.10 & -                & -    & 0.65     &       &            \\
11.&  4831.34313 & 3.24 & -                & -    & 0.21     &       &            \\
12.&  4842.32422 & 1.83 & -                & -    & 0.90     &       &              \\
13.&  4869.29544 & 2.11 & -                & -    & 0.60     &       &             \\
14.&  4870.28783 & 1.71 & -                & -    & 0.66     &       &            \\
15.&  4898.18485 & 1.95 & -                & -    & 0.43     &       &            \\
16.&  4899.28240 & 1.97 & -                & -    & 0.50     &       &            \\
17.&  4959.12810 & 3.53 & -                & -    & 0.28     &       &            \\
\hline
\end{tabular}
\end{scriptsize}
\end{table}

\begin{figure}
\includegraphics[width=9.5cm,height=9.5cm]{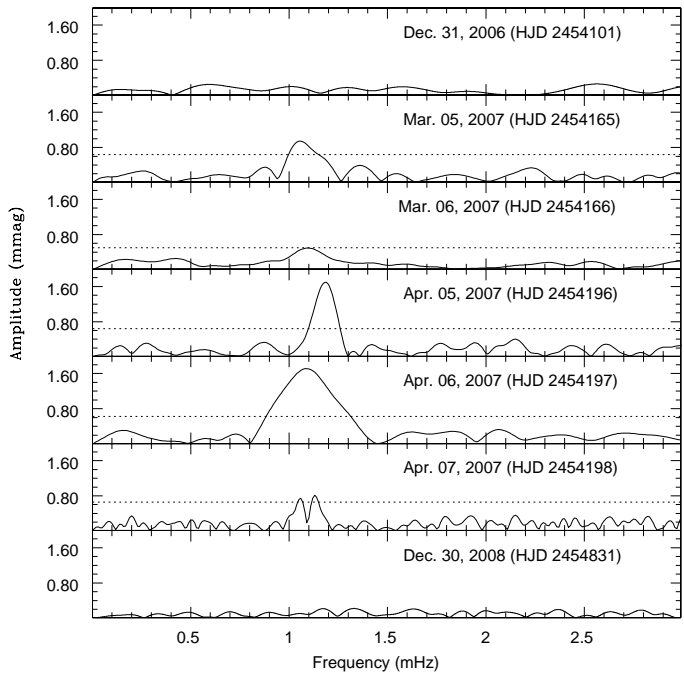}
\caption{The amplitude spectra of the light curves shown in Fig. \ref{light}. The dotted line is   correspond to 98\% confidence level according to criteria of Scargle (1982).}
\label{indivisualft}
\end{figure}

HD103498 was observed photometrically on a total of seventeen nights and Table \ref{log} lists the journal of these observations carried out during the last four years. The quoted error in the frequency is the full width at half-maximum (FWHM) of the peak. The phase ($\phi$, in period fractions) is calculated with the ephemeris used in the construction of the magnetic curve by Auri\`ere~et~al. (\cite{AWS07}) and kindly provided by one of the authors G. Wade:

 HJD(magn. max)= 2450000.0 + 15.830 d

\begin{figure}
\includegraphics[width=9.5cm,height=7.5cm]{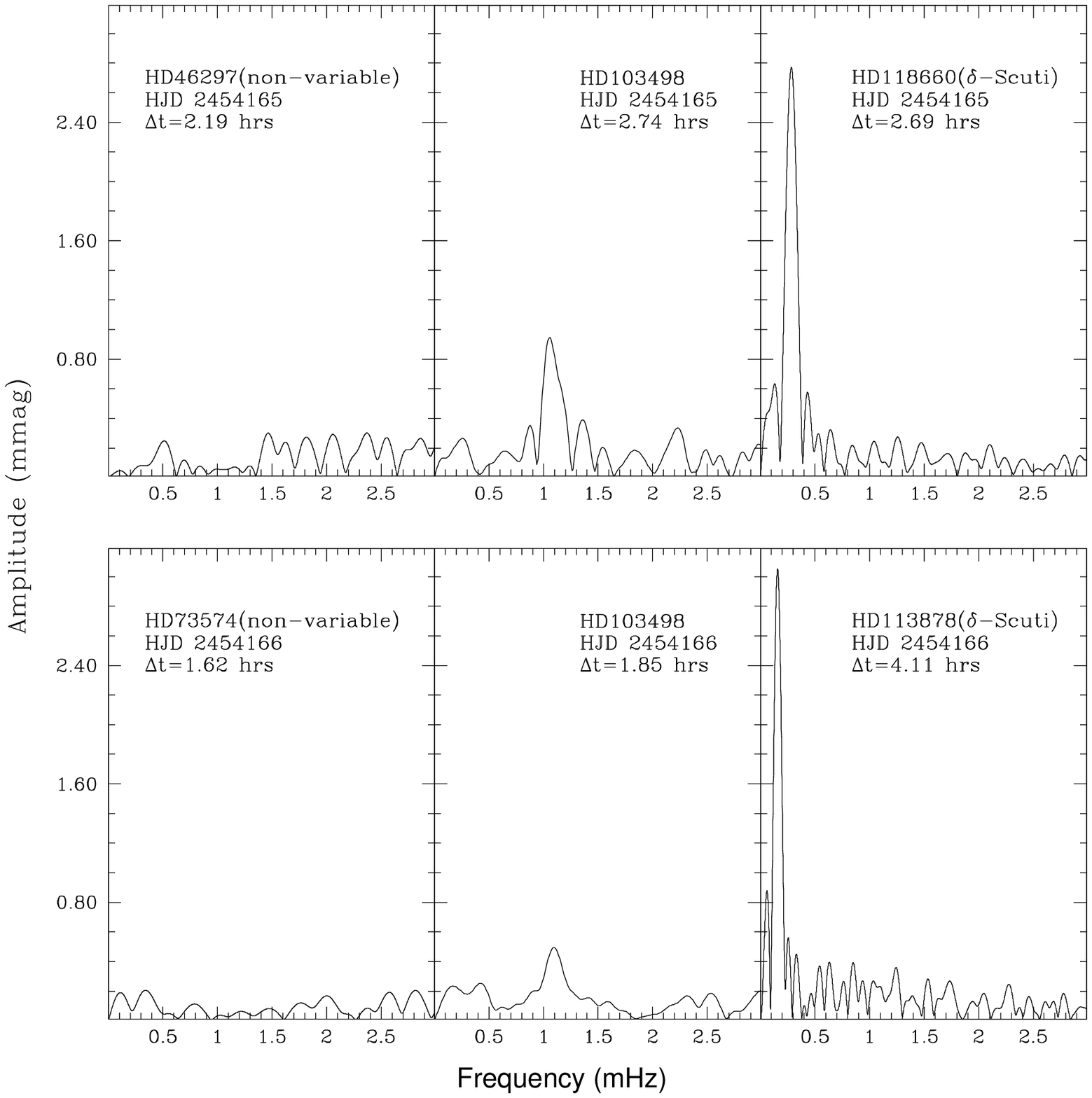}
\caption{Amplitude spectrum of two non-variables and two $\delta$-Scuti stars observed on the same nights when HD\,103498 was observed. The peak at frequency about 1.0-mHz is absent in all the spectra other than HD\,103498. The high-amplitude peaks in the $\delta$-Scuti stars HD\,118660 and HD\,113878 correspond to period range of $\delta$-Scuti stars.}
\label{compft}
\end{figure}

Note that the accuracy of the magnetic period as estimated by Auri\`ere~et~al. (\cite{AWS07}) is $\pm$0.010 d which is not enough to determine the correct phases of the current photometry and spectroscopy. Expected phase errors are within $\pm$0.15-0.20 for the whole data set.  The columns 7 and 8 gives the signal-to-noise ratio and false alarm probability, respectively. Some of the sample light curves of HD\,103498 are shown in Fig. \ref{light} and corresponding amplitude spectra are shown in Fig. \ref{indivisualft}. The horizontal dotted lines in the frequency spectra  correspond to the 98\% confidence level according to  Scargle (\cite{scargle82}) criteria. The long term sky-transparency variations in the time-series data are removed by subtracting a sinusoidal function  of frequency $f_1$, amplitude $A_1$ and phase $\phi_1$ of the form $A_1 \cos(2\pi f_1*t+\phi_1)$ - a technique known as ``prewhitening''. The process was repeated till the sky transparency peaks reduced to the level of scintillation noise. The amplitude spectra (Fig.~\ref{indivisualft}) clearly show that the star light is practically remained constant towards the end of year 2006. A peak appeared at the frequency 1.06-mHz on March 05, 2007 which faded away in the next night. After an interval of a month the amplitude peaked again and subsequently dropped on April 07, 2007. On the same night having the longest observing run, we found a double peak profile. Further observations did not show any prominent peak at this frequency. The frequency spectra also show amplitude modulations which is generally observed in the roAp stars which might be due to one or combined effects of (a) changing aspects as the star rotates, (b) beating between unresolved pulsation modes; and (c) real variations in the amplitude of the mode of pulsations.
Detection of  periodic variability in HD\,103498  is unlikely to be spurious as on two observing nights we also observed two non-variables and
two known $\delta$-Scuti stars just before and after HD\,103498. Figure \ref {compft} shows
the amplitude spectrum  of these stars along with HD\,103498. It may be noticed that the amplitude spectrum for HD\,103498
show a prominent peak at a frequency $\sim$1.1-mHz while the same is absent in others. The prominent peaks in
the amplitude spectrum of HD\,118660 and HD\,113878 correspond to the $\delta$-Scuti pulsation reported by
Joshi et al. (\cite{joshi06}).

To check the significance of the variability seen in the  five nights, Scargle (\cite{scargle82}) ``False Alarm'' criteria
is used to find the false alarm probability F and is expressed as

\begin{equation}
F = 1-[1-exp(-z)]^N
\end{equation}

\noindent
where N is the number of independent  frequencies searched in the time-series. For a given Nyquist frequency
$\nu_N$, N is roughly given as $\cong\nu_N\bigtriangleup$t; where  $\bigtriangleup$t is the total time span
of the data set. The exponential power `z' represents the signal-to-noise ratio in power.  We calculated the false alarm probabilities F for the data sets where the variability was detected. The last column of Table \ref{log} gives the values of F for different data sets. The confidence level of peak on four nights is $>$ 99\% except to April 07, 2007 (HJD2454166) where the level is 97.23\%. Therefore, the variability seen in HD\,103498 has almost no chance of being an artifact.

\section{Spectroscopic analysis}

\subsection{Observations and data reduction}\label{HS-obs}

To check for possible rapid variability of HD\,103498, we performed time-resolved spectroscopy of HD\,103498 using
Fibre-fed Echelle Spectrograph (FIES) at the 2.56-m Nordic Optical Telescope (NOT). The
observations were obtained for a duration of 3.4-hr on the night of February 2, 2009 (HJD
2454865.624--2454865.764) as part of the NOT fast-track Service Mode program
38-415. We collected 71 stellar spectra each of an exposure time of 120-sec.
With the overhead of 48-sec this gave us a sampling rate of approximately one
spectrum every 168-sec, sufficient to resolve the 15-min pulsation period
suspected in HD\,103498. Our observations fall into the magnetic phase range
0.37$\pm$0.2 if the magnetic ephemeris and the error of the period are used.

\begin{figure}
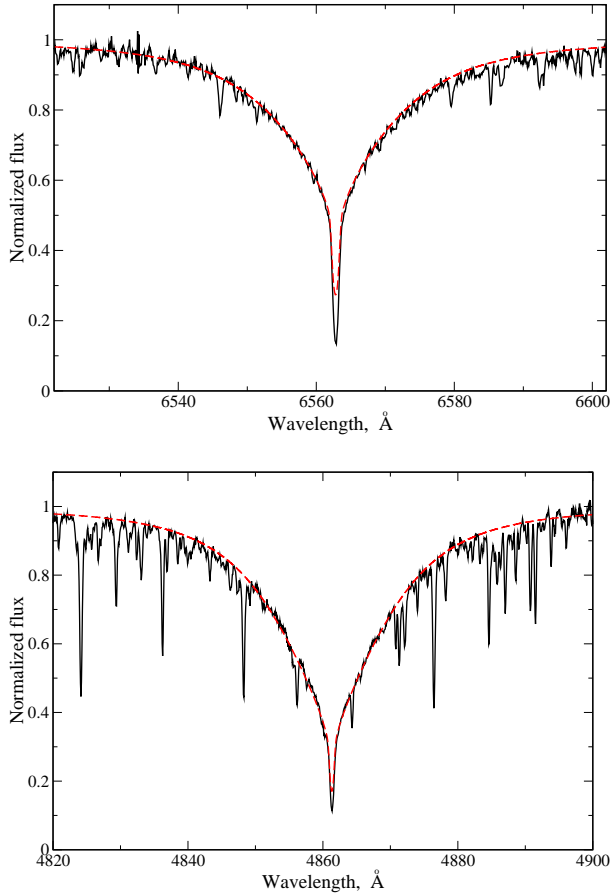

\includegraphics[width=8cm]{H_alpha.eps}

\bigskip

\includegraphics[width=8cm]{H_beta.eps}
\caption{Comparison between the observed and synthesized H$\alpha$ (top) and H$\beta$ (bottom) line profiles for model atmosphere with \teff=9500 K, \lgg=3.6, M=+0.5. }
\label{Hydrogen}
\end{figure}

The FIES instrument was configured to use the medium-resolution mode, which
provides a wavelength coverage of the 3635--7270~\AA\ region at the resolving
power of $R=47000$. We used the REDUCE package of Piskunov \& Valenti
(\cite{PV02}) to perform the standard steps of the echelle spectra reduction
(construction of the master flat field and bias frames, order location,
flat-fielding and wavelength calibration) followed by the optimal extraction of the
stellar spectra. The typical signal-to-noise ratio of the individual
observations is 80--100 around $\lambda$ 5000~\AA. Individual echelle orders of
the extracted spectra were post-processed as described by Kochukhov et
al. (\cite{KRW07}) to ensure a consistency of the continuum normalization for
all 71 spectra.
\begin{table*}
\caption[ ]{LTE atmospheric abundances HD\,103498 with the error
estimates based on the internal scattering from the number of analysed lines,
$n$. Third and forth columns give abundances in similar stars HD\,133792 and HD\,204411 for comparison.
The last column gives the abundances of the solar atmosphere
(Asplund~et~al. \cite{met05}).}
\label{abundance}
\begin{center}
\begin{tabular}{l|cc|c|c|c}
\hline
\hline
Ion &\multicolumn{2}{|c|}{HD\,103498}& HD\,133792  &  HD\,204411 &  Sun \\
    &$\log (N/N_{\rm tot})$ & $n$  &$\log (N/N_{\rm tot})$ & $\log (N/N_{\rm tot})$ &$\log (N/N_{\rm tot})$  \\
\hline
\ion{C}{i}    & ~~$-$4.11$\pm$0.23 &  3 & ~~$-$4.68& ~~$-$4.37 & ~~$-$3.65~ \\			
\ion{O}{i}    & ~~$-$3.92:         &  1 & ~~$-$4.23& ~~$-$4.03 & ~~$-$3.38~ \\			
\ion{Na}{i}   & ~~$-$5.09:         &  1 & ~~$-$5.35& ~~$-$5.28 & ~~$-$5.87~ \\			
\ion{Mg}{i}   & ~~$-$3.58$\pm$0.13 &  4 & ~~$-$3.91& ~~$-$4.34 & ~~$-$4.51~ \\			
\ion{Mg}{ii}  & ~~$-$4.43$\pm$0.27 &  3 & ~~$-$4.18& ~~$-$4.62 & ~~$-$4.51~ \\			
\ion{Al}{ii}  & ~~$-$6.06:         &  1 & ~~$-$6.03& ~~$-$5.85 & ~~$-$5.67~ \\		
\ion{Si}{i}   & ~~$-$3.65$\pm$0.33 &  3 & ~~$-$3.69& ~~$-$4.13 & ~~$-$4.53~ \\		
\ion{Si}{ii}  & ~~$-$3.64$\pm$0.47 &  5 & ~~$-$5.09& ~~$-$4.11 & ~~$-$4.53~ \\			
\ion{Ca}{ii}  & ~~$-$5.91          &  1 & ~~$-$7.36& ~~$-$4.67 & ~~$-$5.73~ \\			
\ion{Ti}{ii}  & ~~$-$6.45$\pm$0.16 & 15 & ~~$-$6.88& ~~$-$6.49 & ~~$-$7.14~ \\			
\ion{V}{ii}   & ~~$-$8.14$\pm$0.18 &  2 & ~~$-$8.14&           & ~~$-$8.04~ \\		
\ion{Cr}{i}   & ~~$-$3.25$\pm$0.23 & 71 & ~~$-$3.79& ~~$-$4.85 & ~~$-$6.40~ \\			
\ion{Cr}{ii}  & ~~$-$3.31$\pm$0.24 &151 & ~~$-$3.75& ~~$-$4.70 & ~~$-$6.40~ \\			
\ion{Mn}{i}   & ~~$-$5.94$\pm$0.11 &  2 & ~~$-$5.55& ~~$-$5.96 & ~~$-$6.65~ \\			
\ion{Mn}{ii}  & ~~$-$5.72$\pm$0.23 &  3 & ~~$-$5.39& ~~$-$5.66 & ~~$-$6.65~ \\			
\ion{Fe}{i}   & ~~$-$2.98$\pm$0.20 & 80 & ~~$-$3.31& ~~$-$3.76 & ~~$-$4.59~ \\			
\ion{Fe}{ii}  & ~~$-$3.01$\pm$0.18 &169 & ~~$-$3.18& ~~$-$3.52 & ~~$-$4.59~ \\			
\ion{Co}{ii}  & ~~$-$5.58:         &  1 & ~~$-$5.99& ~~$-$6.50 & ~~$-$7.12~ \\			
\ion{Ni}{i}   & ~~$-$5.34:         &  1 & ~~$-$6.05& ~~$-$5.68 & ~~$-$5.81~ \\		
\ion{Sr}{ii}  & ~~$\le-$8.5        &  3 & ~~$-$6.36& ~~$-$8.5: & ~~$-$9.12~ \\
\ion{Ba}{ii}  & ~~$-$8.64:         &  1 & ~~$-$8.73& ~~$-$9.02 & ~~$-$9.87~ \\			
\ion{Ce}{ii}  & ~~$-$9.10:         &  1 & ~~$-$9.07:& ~$-$10.26 & ~$-$10.46~ \\		
\ion{Pr}{iii} & ~~$-$8.87$\pm$0.33 &  4 & ~~$-$9.51& ~$<-10.5$ & ~$-$11.33~ \\			
\ion{Nd}{iii} & ~~$-$8.41$\pm$0.16 &  9 & ~~$-$9.08& ~$-$10.05 & ~$-$10.59~ \\			
\ion{Sm}{ii}  & ~~$-$8.70:         &  1 &$\le-$10.4&           & ~$-$11.03~ \\		
\ion{Eu}{ii}  & ~~$-$8.85$\pm$0.12 &  2 & ~~$-$9.81& ~$-$10.95 & ~$-$11.53~ \\
\ion{Gd}{ii}  & ~~$-$8.72:         &  1 & ~~$-$9.60&           & ~$-$10.92~ \\		
\hline											     %
\teff     &\multicolumn{2}{|c|}{9500~K}    & 9400~K    & 8400~K    & 5777~K  \\				
\lgg      &\multicolumn{2}{|c|}{3.6~~~~}   & 3.7~~~~   & 3.5~~~~   & 4.44~~~~\\				
\vt       &\multicolumn{2}{|c|}{1.0~\kms}  &0.0~\kms   &0.0~\kms   & 0.9~\kms\\				
\vs       &\multicolumn{2}{|c|}{12~\kms}   &1.0~\kms   & 6.3~\kms  & 1.9~\kms\\				
$\log(L/L_{\odot})$&\multicolumn{2}{|c|}{2.00}& 2.02   & 2.01      &         \\
\bz, G    &\multicolumn{2}{|c|}{169}       & 120       & 88        &         \\
\hline											
\end{tabular}
\end{center}
\end{table*}
We obtained two ThAr reference spectra, one at the beginning and another at
the end of the stellar time series. Each arc frame was wavelength-calibrated to
the internal precision of 50--60~m\,s$^{-1}$ using $\approx$\,1500 emission lines in
all echelle orders. A RV drift of 120~m\,s$^{-1}$ was measured between the two
ThAr exposures. This change of the spectrograph's zero point, as well as a small
change of the heliocentric RV that occurred during our observations, was
compensated by subtracting a straight line fit from the RV measurements of the
individual lines and spectral regions.

In addition to FIES data we also used one of the MuSiCoS spectra employed for magnetic field measurements
(Auri\`ere~et~al. \cite{AWS07}). This spectrum, kindly provided to us by G. Wade, was
obtained at JD=2452255.73 close to the phase of magnetic minimum with the signal-to-noise ratio about 200,
and with the spectral resolution of $R=35000$. The details of general data reduction are given by Auri\`ere~et~al. (\cite{AWS07}).
For the purpose of our analysis we slightly re-normalized the continuum.

\subsection{Abundance analysis}\label{analysis}

We performed full spectroscopic analysis of HD\,103498 to identify its evolutionary status and the 
chemical peculiarities.
Abundance study of the star was based primarily on the averaged spectrum of our time-series spectroscopy in the spectral
region 3900--7270 \AA. We also repeated abundance analysis using the MuSiCoS spectrum and obtained results in agreement
with those derived from the averaged FIES spectrum within the errors of abundance determination. For hydrogen line profiles we used
the MuSiCoS spectrum where continuum is better defined in the region of the line wings.

\subsubsection{Fundamental parameters}\label{atm}

Atmospheric parameters of HD\,103498 were estimated using calibration of Str\"omgren (Moon \& Dworetsky \cite{MD})
and Geneva photometry (Kunzli~et~al. \cite{Geneva}) realized in the package TEMPLOGG (Kaiser \cite{TMG}).
Photometric indices were taken from Hauck \& Mermilliod (\cite{Hauck_M}, Str\"omgren) and from Rufener (\cite{Rufener}, Geneva).
An average effective temperature \teff=9370$\pm$140 and \lgg=3.90 were obtained. Analysis of H$\alpha$ and H$\beta$ line profiles
shows that while effective temperature estimate is reliable, surface gravity is certainly overestimated.
A reasonable fit of the calculated H$\alpha$ and H$\beta$ line profiles to the observed ones was obtained with
the following atmospheric parameters: \teff=9500~K, \lgg=3.6 (Fig.\,\ref{Hydrogen}). We used these parameters in all
further abundance calculations.

To estimate the luminosity we did not use parallax value 3.37 mas of the star as given by
van Leeuwen (\cite{Lwn07}) for an individual object. If HD\,103498 (65~UMa~D) and 65~UMa~AC comprise a
physical pair (Pourbaix et al. \cite{tokovinin}), then both stars should have similar parallaxes. Therefore,
we adopted a weighted mean parallax value of the entire system, 4.02$\pm$0.40 mas, for HD\,103498.
Reddening parameter $E(B-V)$\,=\,0.007 and interstellar absorption $A_v$\,=\,0.021~mag were estimated from the UBV photometric data of Huchra \& Willner (\cite{HW73}).
The derived luminosity of the star, $\log(L/L_{\odot})$=2.00$\pm$0.09,
is very close to the value $\log(L/L_{\odot})$=2.06 determined by Auri\`ere~et~al. (\cite{AWS07}).
Fundamental parameters of HD\,103498 as well as its chemical peculiarities (see next Section) are very
similar to other evolved Ap stars HD\,133792 (Kochukhov~et~al. \cite{vip}) and HD\,204411 (Ryabchikova~et~al. \cite{RLK05}).
Position of HD~103498 on HR diagram (see Fig.\,\ref{HR}) provides an additional support for the lower gravity derived from
hydrogen line profiles compared to photometric calibrations.

\subsubsection{Atmospheric abundances}\label{abun}
\begin{figure*}
\resizebox{\hsize}{!}{\rotatebox{0}{\includegraphics{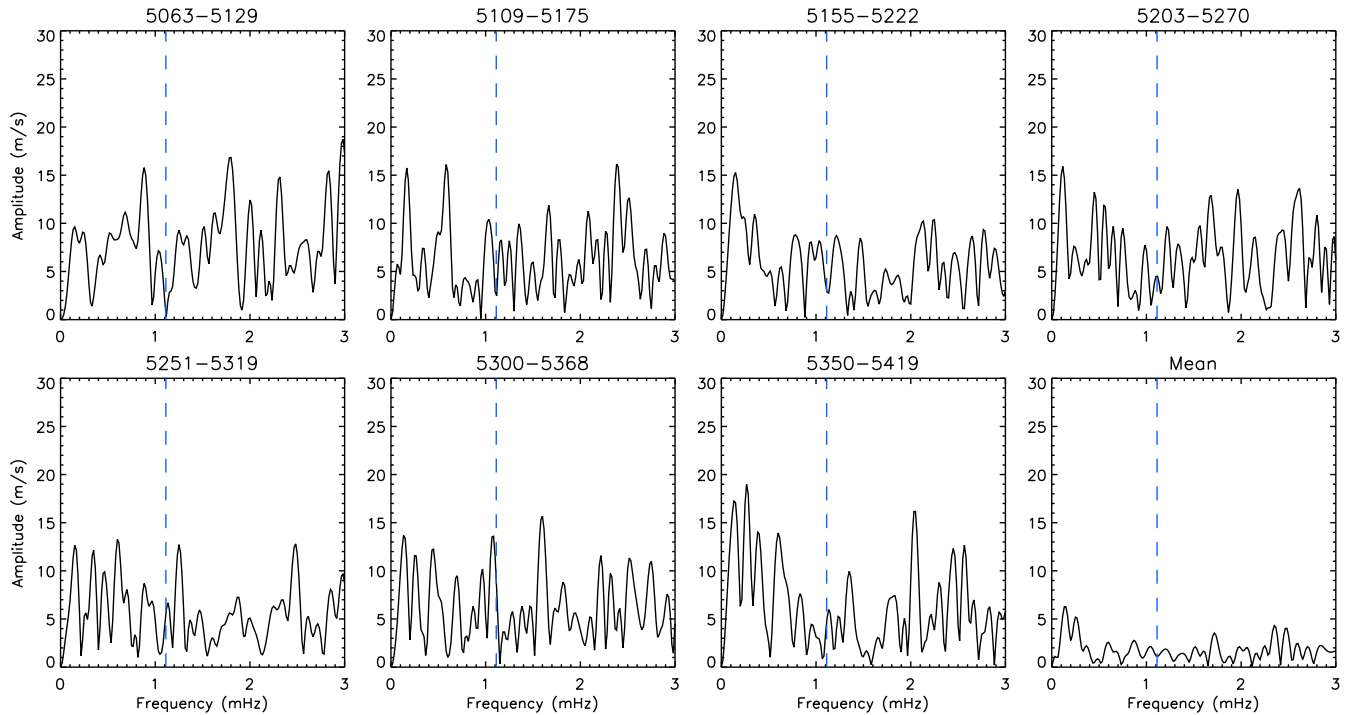}}}
\caption{Amplitude spectra of the cross-correlation RV measurements for 7
representative echelle orders of the FIES spectra of HD\,103498.
The lower right panel shows the amplitude spectrum
for velocity obtained by combining RV curves of 16 best orders.
The vertical dashed line indicates mean frequency identified in
photometric observations.}
\label{ft_cor}
\end{figure*}

Identification of lines in the spectrum of HD\,103498 was
based on the theoretical spectrum calculated for the entire observed spectral region
using the line extraction from \vald\ (Kupka et al. \cite{VALD2} and references
therein) and {\sc DREAM} (Bi\'emont et al. \cite{DREAM}) databases. Atomic data on the rare-earth elements (REE)
compiled in  the DREAM database were extracted via the VALD interface. Comparison of the
synthetic and  observed spectra allowed us to choose the least blended lines for the abundance
analysis. Direct comparison of HD\,103498 spectrum with spectrum of well-studied Ap star
HD\,133792 (Kochukhov~et~al. \cite{vip}) shows a pronounced similarity between two objects: both spectra are
crowded with strong lines of Fe-peak elements while the lines of the REE are weak or absent. \ion{Cr}{ii}
and \ion{Fe}{ii} lines with the lower level excitation energy above 10 eV are numerous and many of them are rather strong,
which is an indication of the extreme Cr and Fe overabundances in stellar atmosphere.

LTE abundance determination is based mainly on the equivalent widths
analysed with the improved version of \width\ code updated for the \vald\
output line lists and kindly provided to us by V. Tsymbal. Model atmosphere was calculated with the
\atlas\ code (Kurucz \cite{K93}) using a metallicity $[M/H]=+0.5$. To make a proper selection of unblended and
minimally blended lines for abundance calculations we synthesized the whole observed spectral region
3965--7270~\AA\ with the help of \synth\ code (Kochukhov \cite{synth3}). The best fit to the observed
unblended line profiles was achieved for \vs\,=\,12~\kms, in close agreement with \vs\,=\,13~\kms\ derived by Auri\`ere~et~al.
(\cite{AWS07}) from the mean metal line profile. Microturbulent velocity, \vt\,=\,1.0$\pm$0.2~\kms, was obtained as an averaged value
between the values derived from numerous lines of \ion{Cr}{i}, \ion{Cr}{ii}, \ion{Fe}{i}, and \ion{Fe}{ii}.
Table~\ref{abundance} summarizes the results of our abundance analysis. For each species an average abundance with rms
based on 'n' measured lines is given. For comparison we also give abundances in the atmospheres of two
CrFe-rich evolved Ap stars HD\,133792 (Kochukhov~et~al. \cite{vip}) and HD\,204411 (Ryabchikova~et~al. \cite{RLK05}) and
in the solar photosphere (Asplund~et~al. \cite{met05}).

The results show that HD\,103498 have the largest Cr and Fe overabundances among the three stars.
It is also Si-overabundant, but silicon seems to be stratified similar to HD\,133792 and HD\,204411.
Calcium lines are weak and we analysed only one \ion{Ca}{ii} line  $\lambda$~6456.88~\AA. Trying to fit the
resonance \ion{Ca}{ii} $\lambda$~3933.66~\AA\ line we found that description of the line core requires 10 times smaller abundance than
the wings and the 6456.88~\AA\ line. This indicates the presence of Ca stratification similar to the one in HD\,133792
(Kochukhov~et~al. \cite{vip}) and in HD 204411 (Ryabchikova~et~al. \cite{RLK05}). We also found that the core of
\ion{Ca}{ii}~$\lambda$~3933.66 is blue-shifted by $\sim$0.02~\AA. Blue shift of this line was also found in HD\,133792
which is known to have strong Ca isotopic anomaly (see Ryabchikova~et~al. \cite{RKB08}, Section 6). To investigate the Ca isotopic
anomaly in HD\,103498 one needs spectral
observations in the region of IR \ion{Ca}{ii} triplet that has the largest isotopic separation.

Spectral synthesis of the regions around resonance \ion{Sr}{ii}~$\lambda$~4077, 4215~\AA\ as well as \ion{Sr}{ii}~$\lambda$~4305~\AA\
lines clearly shows that lines of Cr and Fe, not of \ion{Sr}{ii}, provide the main contribution to the observed strong features appearing at
the position of the \ion{Sr}{ii} lines. In fact, the Sr abundance in HD\,103498 does not exceeded much the solar one. Thus, classification of the Cr-rich stars as Cr-Sr based on low-resolution spectra may be incorrect. Furthermore, it is interesting to note that the two stars with substantial rotation, HD\,103498
and HD\,204411, have the same small Sr overabundance while HD\,133792 with negligible rotation is very Sr-overabundant but has similar abundances of most other elements.

\subsection{Cross-correlation radial velocity analysis}

Due to a substantial difference of the atmospheric parameters and spectral
appearance of HD\,103498 compared to the known roAp stars, we do not have an
\textit{a priori} information which lines might be most suitable for detection of
the velocity oscillations. In this situation a cross-correlation analysis of large
spectral regions appears to be the optimal approach to the spectroscopic search of
pulsations. In roAp stars metal lines exhibit a large scatter in the pulsation
amplitude and phase, leading to a significant dilution of the signal coming from
strongly pulsating lines by lines showing weak of no variation. However, even in
this case cross-correlation studies are usually successful in detecting reliably
oscillation signals with amplitudes as low as few m\,s$^{-1}$ (e.g., Hatzes \&
Mkrtichian \cite{HM04}).

We applied the cross-correlation analysis to the individual echelle orders of the
FIES spectra of HD\,103498. The time-resolved spectra were compared to the mean one
using the chi-square criterion and the RV shift minimizing the chi-square was
recorded. Only the spectral regions containing absorption features deeper than 3\%
were included in the analysis. The RV curves obtained with this method were analysed
using the standard discrete Fourier transform (FT) technique. Fig.~\ref{ft_cor}
shows the amplitude spectra for 7 consecutive echelle orders in the spectral region
characterized by a high $S/N$ and a large line density. No statistically significant
periodicity is evident. The highest noise peaks reach 15--20~m\,s$^{-1}$. We also
analysed a mean RV curve obtained by  combining the measurements for 16 echelle
orders for which the best RV precision was achieved. The corresponding FT spectrum
is shown in the lower right panel of Fig.~\ref{ft_cor}. It reveals no oscillations
stronger than $\approx5$~m\,s$^{-1}$. The upper limit for the pulsations in the
vicinity of the photometric frequency of HD\,103498 is only 2.8~m\,s$^{-1}$.

\begin{figure}
\resizebox{8cm}{!}{\rotatebox{0}{\includegraphics{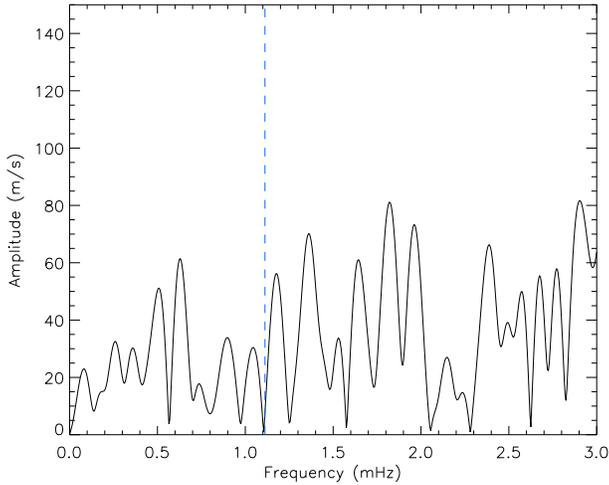}}}
\caption{Amplitude spectrum of the center-of-gravity RV measurements averaged
for 8 lines of \ion{Pr}{iii} and \ion{Nd}{iii} in HD\,103498. The vertical dashed line indicates mean frequency identified in
photometric observations.}
\label{ft_REE}
\end{figure}

In most of roAp stars the lines of the REE in the first and second ionization stages usually
show the highest RV amplitudes. In HD\,103498 these lines are either practically absent (\ion{REE}{ii}) or
rather weak (\ion{REE}{iii}), while the lines of Fe-peak elements are numerous and strong. Therefore a pulsation signal from the REE lines may be lost in cross-correlation analysis. We made center-of-gravity RV measurements of the individual lines of \ion{Pr}{iii}~$\lambda\lambda$~5300, 6867~\AA\ and  \ion{Nd}{iii}~$\lambda\lambda$~4914, 4927, 5050, 5102, 5294, 6327~\AA\ and analysed resulting mean RV curve for the two REE species. The corresponding FT spectrum, shown in Fig.~\ref{ft_REE}, provides no evidence of the RV pulsation signal in the vicinity of the photometric frequency. However, the upper limit of the pulsation amplitude is fairly large, $\approx$80~\ms.

\section{Discussions}

The  detection of 15-min periodic oscillations in the photometric data of year 2007 followed by non-detection in the spectroscopic data of year 2009 are of vast importance. The possible reason for this phenomena may be significant intrinsic changes of the stellar pulsational amplitude. The  absence of periodic light variations in the photometric data sets  carried out during  2006,2008 and 2009 could be due to the fact  that the real amplitude variations might be buried under the noise. The detection limit of the photometric variability depends on the atmospheric noise which consists of the scintillation noise  and the long term sky transparency variations.
\begin{figure}
\includegraphics[width=8.5cm,height=7.0cm]{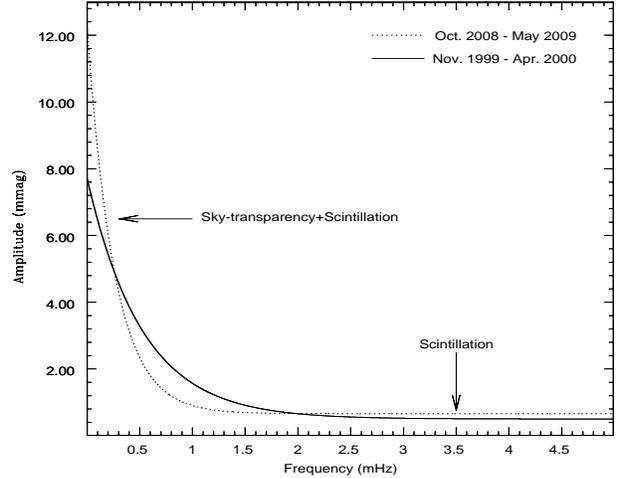}
\caption{The average sky-transparency and scintillation noise obtained by analysis of high-speed photometric observations of  year 1999-2000 (solid line) and 2008-09 (dotted line). The scintillation noise appears at higher frequencies ($>$ 1-mHz) region which sets  the limit of detection of high-overtone pulsations (roAp like oscillations).  The long term sky transparency variations lie in lower frequencies ($<$ 1-mHz)  region where it is very difficult to distinguish the low-overtone pulsations ($\delta$-Scuti type)  with  the sky transparency variations.}
\label{skytransparancy}
\end{figure}
Fig. \ref{skytransparancy} shows the average atmospheric noise  at ARIES observatory which is based on the high-speed photometric observations carried out under  the ``Nainital-Cape Survey''  project aiming to search and study  the pulsational variability in chemically peculiar stars  (Martinez et al. \cite{martinez01}; Joshi et al. \cite{joshi06}; Joshi et al. \cite{joshi09}). The fitted solid line corresponds to the Fourier transform of  the light curves of 47 stars observed on 84 occasions during 1999-2000 and the dotted line is for 21 stars observed on 33 occasions during 2008-2009. The respective amplitude of the sky transparency variations and scintillation noise in year 1999-2000 are 7.23 and 0.50-mmag and the corresponding values for year 2008-2009 are 11.68 and 0.66-mmag, respectively.  This clearly shows that the atmospheric noise has been increased slightly in last ten years owing to enhanced human activities around the observatory.  The atmospheric noise can be minimized by installing bigger telescopes at a good observing site where the sky is stable and photometric (Young \cite{young67}).  Towards this, ARIES is now in the process of installing 1.3-m and 3.6-m optical telescopes at a new astronomical site, Devasthal (longitude: $79^{o}40^{'}57{''} $ E, latitude : $29^{o}22{'}26{'}' $ N, altitude : 2420-m) by the end of year 2009 and 2012, respectively. This site has an average seeing of $\approx$ 1$^{\prime\prime}$ near the ground and $\approx$ 0.65$^{\prime\prime}$ at 12-m above the ground (Sagar et al. \cite{sagar00}; Stalin et al. \cite{stalin01}). In the near future,  the new observing facilities at Devasthal shall contribute significantly in  the area of asteroseismology.

It is possible to attribute the non-detection of pulsations in the spectroscopy to the observation at unfavorable rotation phase, $\varphi\approx0.25$ or 0.75 when the magnetic equator passes in front of the observer. Unfortunately, the magnetic ephemeris of HD\,103498 is not known with enough accuracy to establish the precise rotational phase of our observations. On the other hand, we do not
observe clear modulation of the photometric pulsational amplitude on the time scale of the rotation period which argues against such modulation in spectroscopy. Furthermore, spectroscopic studies of a few roAp stars which were followed over the entire rotation cycle (e.g., Kochukhov \cite{K04}; Mkrtichian \& Hatzes \cite{MH05}) show that RV pulsation never disappears entirely.

\begin{figure}
\includegraphics[width=9cm,height=9cm]{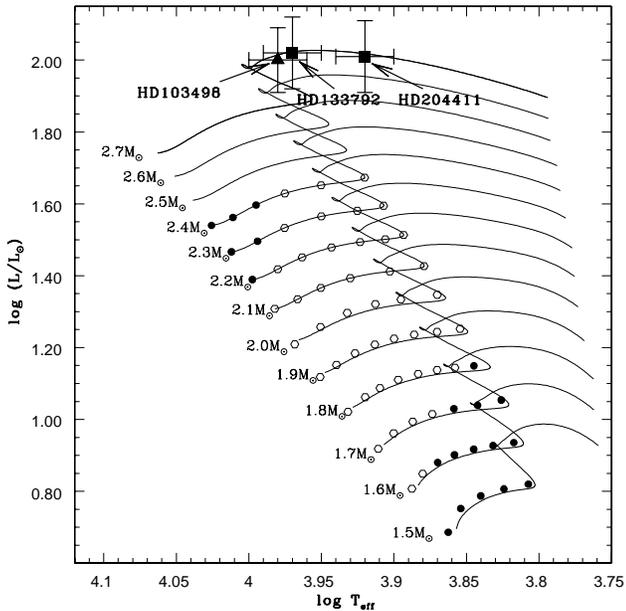}
\caption{HR diagram showing the position of HD\,103498 (by symbol $\blacktriangle$) having \teff=9500 $\pm$200K and $\log(L/L_{\odot})=2.0$. For comparison, positions of two evolved Ap stars HD\,133792 and  HD\,204411 having \teff=9400$\pm$190 K and $\log(L/L_{\odot})=2.02$; and  \teff=8400$\pm$200 K and $\log(L/L_{\odot})=2.01$, respectively, are also shown (by symbol $\blacksquare$). 
The filled circles show radiative envelope models in which no high-order acoustic oscillations were found, while open circles show the models in which the latter were found.
}
\label{HR}
\end{figure}

Fig.\,\ref{HR} shows the position of HD\,103498  on the H-R diagram where an instability strip for roAp stars is
indicated (Cunha \cite{cunha02}). For comparison positions of the two evolved Ap stars, HD\,133792 and HD\,204411, are shown. The stellar evolutionary tracks for the mass range from 1.5 to 2.7 $M_\odot$ (Christensen-Dalsgaard \cite{christ93}) are also over-plotted. Pulsation calculations by Cunha (\cite{cunha02}) and a more recent theoretical study by Th\'eado et al. (\cite{TDNF09}) predict the excitation of pulsations in relatively hot and more evolved Ap stars. However, all of about 40 currently known roAp stars have temperatures in the \teff\ interval from 6400 to 8100 K. The search for RV pulsations in evolved stars with \teff\ larger than 8100~K was not successful (Freyhammer~et~al. \cite{FKC08}). If confirmed the discovery of pulsations in HD\,103498 with  \teff=9500~K may have a profound consequences for our understanding of the excitation of p-modes in magnetic Ap stars.

\section{Conclusions}

The time-series photometric observations of HD\,103498 taken on five nights show the clear pulsational variability of about 15-min with amplitude modulation. However, the same periodicity could not be confirmed in the follow-up photometric and high-resolution spectroscopic observations. Hence, the pulational variability in HD\,103498 should be taken in account very cautiously. The abundance
analysis of the HD\,103498 shows that the lines of the rare-earth elements -- the best indicators of RV pulsations in roAp stars -- are rather
weak. It is also observed that the core of \ion{Ca}{ii}~$\lambda$~3933.66 is blue-shifted by $\sim$0.02~\AA, which may be a result of anomalous Ca isotopic composition. The cross-correlation RV analysis revealed no oscillations with the amplitudes above
$\approx$\,5-m\,s$^{-1}$.  The measurements of the center-of-gravity of eight individual REE lines also did not show any evidence of RV pulsation signal in the vicinity of the photometric frequency. More time-series photometric and spectroscopic observations are required to study the variable nature of  HD\,103498.

\section{Acknowledgments}
The authors are thankful to Prof. Ram Sagar for encouraging to initiate the Indo-Russian collaboration. 
We thank Dr. G. Wade for providing us the MuSiCoS spectra of HD\,103498 and the NOT telescope staff for 
helping in collecting the spectra of HD\,103498. We thank to reviewer Dr. Pierre North for the useful comments and 
suggestions which lead the significant improvement of the manuscript. Resources provided by the electronic databases 
(VALD, SIMBAD,NASA's ADS) are acknowledged. This work was supported by the Presidium RAS program, by research grant 
from the RFBI (08-02-00469a),  by Russian Federal Agency on Science and Innovation
(02.740.11.0247), INT/ILTP/B-3.19 and INT/ILTP/B-3.16. O.K. is a Royal Swedish Academy of Sciences Research Fellow supported by a grant from the Knut and Alice Wallenberg Foundation. Part of this work was carried out under the Indo-South Africa Science and Technology Cooperation (INT/SAFR/P-04/2002/13-02-2003 and INT/SAFR/P-3(3)2009) funded by Departments of Science and Technology of the Indian and South African governments. SJ acknowledges his collegues for the critical reading of the manuscript.


\begin{thebibliography}{}
\bibitem[1995]{AM95}Abt H. A., Morrell N. I., 1995, \apjs, 99, 135
\bibitem[2001]{ashoka01}Ashoka, B. N., Kumar., Babu, V. C., et al. 2001, JAA, 22, 131
\bibitem[2005]{met05}Asplund M., Grevesse N., Sauval A.~J., 2005, ASP Conf. Ser., 336, 25
\bibitem[2007]{AWS07}Auri\`ere M., Wade G., Silvester J., et al., 2007, \aap, 475, 1053
\bibitem[1999]{DREAM}Bi\'emont E., Palmeri P., Quinet P., 1999, \apss, 269-270, 635
\bibitem[2003]{bychkov03}Bychkov V. D., Bychkova L. V., Madej, 2003, \aap, 407, 631
\bibitem[1993]{christ93}Christensen-Dalsgaard J., 1993, Baglin A., Weiss W. W., eds, Proc. IAU Coll. 137, Inside the stars, ASP Conf. Ser. Vol. 40, 483
\bibitem[2002]{cunha02}Cunha M., 2002, \mnras, 333, 47
erratum PASP, 110, 1118 (1998)
\bibitem[1975]{deeming75}Deeming T. J., 1975, Ap\&SS, 36, 137
\bibitem[2008]{FKC08}Freyhammer L. M., Kurtz D. W., et al., 2008, \mnras, 385, 1402
\bibitem[2004]{HM04}Hatzes A. P., Mkrtichian D. E., 2004, \mnras, 351, 663
\bibitem[1998]{Hauck_M}Hauck B. \& Mermilliod M. 1998,  \aaps, 129, 431
\bibitem[1973]{HW73}Huchra J., Willner S. P., 1973, PASP, 85, 85
\bibitem[2006]{joshi06}Joshi S., Mary D. L. et al., 2006, \aap, 455, 303
\bibitem[2009]{joshi09}Joshi S., Mary D. L.,  et al., 2009, in press,  \aap, arXiv0909.0810
\bibitem[2006]{TMG}Kaiser A., 2006, in Astrophysics of Variable Stars, eds. C. Sterken, and C. Aerts, ASP Conf. Ser., 349, 257
\bibitem[2009]{HD75445}Kochukhov O., Bagnulo S., Lo Curto G., Ryabchikova T., 2009, \aap, 493, L45
\bibitem[2007]{synth3}Kochukhov O. 2007, in Physics of Magnetic Stars, eds. D.O.~Kudryavtsev and I.I~Romanyuk, Nizhnij Arkhyz., p.109
\bibitem[2007]{KRW07}Kochukhov O., Ryabchikova T., Weiss W. W., Landstreet J. D., Lyashko D., 2007, \mnras, 376, 651
\bibitem[2006]{vip}Kochukhov O., Tsymbal V., Ryabchikova T., Makaganyk V., Bagnulo S., 2006, \aap, 460, 831
\bibitem[2004]{K04}Kochukhov, O. 2004, \aap, 446, 1051
\bibitem[1997]{Geneva}Kunzli M., North P., Kurucz R. L., Nicolet B., 1997, \aaps, 122, 51
\bibitem[1999]{VALD2}Kupka F., Piskunov N., Ryabchikova T. A., Stempels, H. C., Weiss, W. W., 1999, \aaps, 138, 119
\bibitem[1993]{K93}Kurucz, R. L., 1993, Kurucz CD-ROM 13, Cambridge, SAO
\bibitem[2001]{martinez01}Martinez, P., Kurtz, D. W.,  et al. 2001, A\&A, 371, 1048\
\bibitem[2005]{MH05}Mkrtichian, D. E., Hatzes, A. P. 2005, \aap, 430, 263
\bibitem[1985]{MD}Moon T. T., Dworetsky M. M., 1985, MNRAS, 217, 305
\bibitem[2002]{PV02}Piskunov N. E., Valenti J. A., 2002, \aap, 385, 1095
\bibitem[2004]{tokovinin}Pourbaix D., Tokovinin A. A., Batten A. H., et al., 2004, \aap, 424, 727
\bibitem[1976]{Rufener}Runefer F., 1976, \aaps, 26, 275
\bibitem[2008]{RKB08}Ryabchikova T., Kochukhov O., Bagnulo S., 2008, \aap, 480, 811
\bibitem[2005]{RLK05}Ryabchikova T., Leone F., Kochukhov O., 2005, \aap, 438, 973
\bibitem[1999]{sagar99}Sagar R., 1999, Current Science, Vol. 77, No. 5, p. 643
\bibitem[2000]{sagar00} Sagar, R., Stalin, C. S., et al. 2000, A\&AS, 144, 349
\bibitem[1982]{scargle82}Scargle J. D., 1982, \apj, 263, 835
\bibitem[2001]{stalin01}Stalin, C. S., Sagar, R., et al., 2001, BASI, 2001, 29, 39
\bibitem[2009]{TDNF09}Th\'eado S. Dupret M.-A., Noels A., Ferguson J. W., 2009, \aap, 493, 159
\bibitem[2007]{tiwari07}Tiwari S. K., Chaubey U. S., Pandey C. P., 2007, IBVS, 5900
\bibitem[2007]{Lwn07}van Leeuwen, F, 2007, \aap, 474, 653
\bibitem[1967]{young67} Young A. T., 1967, AJ, 72, 747
\end{thebibliography}
\end{document}